\documentclass[epj]{webofc}
\usepackage[varg]{txfonts}   
\usepackage{color}

\woctitle{Connecting The Dots 2016}

\begin{document}
\title{Online Reconstruction and Calibration with feed back loop in the ALICE High Level Trigger}

\author{\firstname{David} \lastname{Rohr}\inst{1}\fnsep\thanks{\email{drohr@jwdt.org}} \and
        \firstname{Ruben} \lastname{Shahoyan}\inst{2}\fnsep\thanks{\email{ruben.shahoyan@cern.ch}} \and
        \firstname{Chiara} \lastname{Zampolli}\inst{2,3}\fnsep\thanks{\email{Chiara.Zampolli@cern.ch}} \and
        \firstname{Mikolaj} \lastname{Krzewicki}\inst{1,4}\fnsep\thanks{\email{mikolaj.krzewicki@cern.ch}} \and
        \firstname{Jens} \lastname{Wiechula}\inst{4}\fnsep\thanks{\email{Jens.Wiechula@cern.ch}} \and
        \firstname{Sergey} \lastname{Gorbunov}\inst{1,4}\fnsep\thanks{\email{Sergey.Gorbunov@cern.ch}} \and
        \firstname{Alex} \lastname{Chauvin}\inst{5}\fnsep\thanks{\email{alex.chauvin@cern.ch}} \and
        \firstname{Kai} \lastname{Schweda}\inst{6}\fnsep\thanks{\email{kschweda@physi.uni-heidelberg.de}} \and
        \firstname{Volker} \lastname{Lindenstruth}\inst{1,4}\fnsep\thanks{\email{voli@compeng.de}} for the ALICE Collaboration
}

\institute{Frankfurt Institute for Advanced Studies, Ruth-Moufang-Str.~1, 60638 Frankfurt \and
           European Organization for Nuclear Research (CERN), Geneva, Switzerland \and
           Sezione INFN, Bologna, Italy \and
           Goethe University Frankfurt \and
           Technical University of Munich \and
           University of Heidelberg
          }

\abstract{%
ALICE (A Large Heavy Ion Experiment) is one of the four large scale experiments at the Large Hadron Collider (LHC) at CERN.
The High Level Trigger (HLT) is an online computing farm, which reconstructs events recorded by the ALICE detector in real-time.
The most compute-intense task is the reconstruction of the particle trajectories.
The main tracking devices in ALICE are the Time Projection Chamber (TPC) and the Inner Tracking System (ITS).
The HLT uses a fast GPU-accelerated algorithm for the TPC tracking based on the Cellular Automaton principle and the Kalman filter.
ALICE employs gaseous subdetectors which are sensitive to environmental conditions such as ambient pressure and temperature and the TPC is one of these.
A precise reconstruction of particle trajectories requires the calibration of these detectors.
As first topic, we present some recent optimizations to our GPU-based TPC tracking using the new GPU models we employ for the ongoing and upcoming data taking period at LHC.
We also show our new approach for fast ITS standalone tracking.
As second topic, we present improvements to the HLT for facilitating online reconstruction including a new flat data model and a new data flow chain.
The calibration output is fed back to the reconstruction components of the HLT via a feedback loop.
We conclude with an analysis of a first online calibration test under real conditions during the Pb-Pb run in November 2015, which was based on these new features.
}
\maketitle
\section{Introduction}
\label{intro}

\subsection{The ALICE Detector}

ALICE (A Large Heavy Ion Experiment) is one of the four major experiments at the Large Hadron Collider (LHC) at CERN in Geneva~\cite{bib:alice}.
While the other large experiments focus mainly on proton-proton collisions, the main purpose of ALICE is to study heavy ion collisions.
This enables the investigation of matter under extreme conditions of high temperature and pressure.
In ion physics mode, the LHC collides lead nuclei at an interaction rate of around $8$\,kHz.
ALICE employs several detectors to measure particle trajectories, energy deposition, and to identify particles.

\subsection{The High Level Trigger (HLT)}

The High Level Trigger (HLT)~\cite{bib:hlt} is an online computing farm consisting of around~$200$ compute nodes for the online processing of the collisions recorded by ALICE.
In contrast to the posterior long-running offline physics analysis, the HLT performs the first processing and analysis in real time.
This involves the reconstruction of the events, calibration of the detectors, data compression to reduce the amount of data stored to tape, online QA, as well as triggering the readout or tagging of physically interesting events.
The HLT receives the data from the experiment via several hundred optical links.
Inside the HLT, independent processing components perform individual steps of the reconstruction and processing.
A custom data transport framework transfers the data between the processing components on the same servers via shared memory or on different servers via an Infiniband network.
The maximum possible data input rate over the detector links is above 60\,GB/s while in normal operation the HLT receives up to 30\,GB/s of recorded data.
The HLT is capable of full real time event reconstruction of the data recorded by ALICE.
The computationally most intensive step is the reconstruction of particle trajectories, also called tracking.

\subsection{Online reconstruction and online calibration}

Several of the ALICE subdetectors are sensitive to environmental conditions such as ambient pressure and temperature.
Precise reconstruction of particle trajectories requires the calibration of these detectors.
Since the environmental conditions change during data taking, calibration must be performed regularly as a single calibration step at the beginning of a run is insufficient.
Performing the detector calibration (or a part thereof) online in the HLT has several advantages:
\begin{itemize}
\item If the calibration result is made available to the online reconstruction in the HLT, this can significantly improve the quality of online reconstruction.
\item Performing the calibration while the data is recorded allows an immediate and better QA (Quality Assurance) already during data taking.
\item Online calibration can render certain offline calibration steps obsolete possibly reducing the computational load during offline reconstruction.
\item Looking ahead to future experiments like ALICE in LHC Run 3 or FAIR at GSI~\cite{bib:fair}, data compression will rely on reconstruction which makes online calibration a necessity.
\end{itemize}

\section{Approach for online calibration}

On one hand, the ALICE calibration is based on reconstruction results like particle trajectories;
on the other hand, the calibration results should be used to improve the reconstruction.
This imposes a cyclic dependency between reconstruction and calibration.
On top of that, calibration involves long running tasks that can last many seconds if not minutes.
This makes it impossible to apply the calibration result to the reconstruction of the events that are used for the calibration.
That would require caching of data for a considerable amount of time which is not possible in the HLT.

\begin{figure}[b]
\centering
\includegraphics[width=\textwidth]{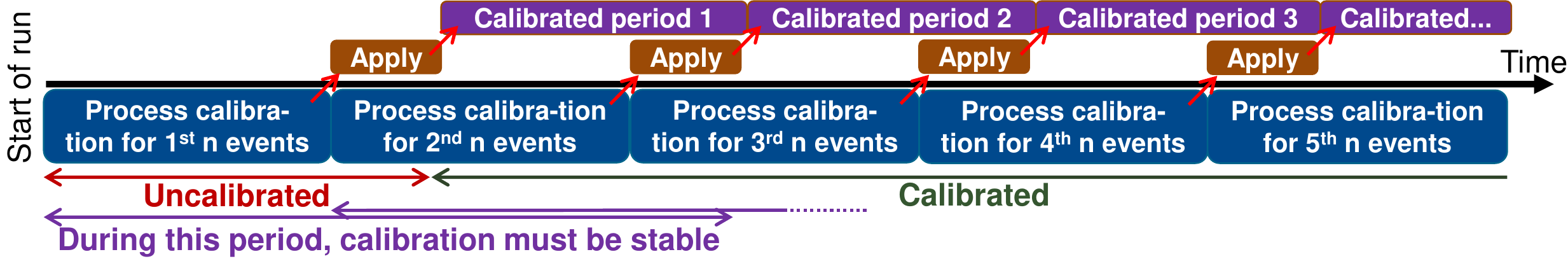}
\caption{Illustration of the approach for online calibration.
The blue boxes are the intervals where the calibration data is aggregated.
Afterwards there is a short delay to prepare new TPC transformation maps based on the calibration and distribute them in the cluster (brown box).
Finally, the HLT reconstruction uses the calibration as long as it is stable (purple box).}
\label{fig:plot-calib-scheme}
\end{figure}

Instead, the HLT employs a different approach for online calibration.
The ambient conditions that affect the calibration are stable over a certain time period.
Even in case of a sudden, total weather change, pressure and temperature will change smoothly such that the calibration created for a certain point in time is valid for a certain period.
In the following, we will discuss the TPC drift velocity as an example.
In this case, the calibration is assumed to be valid for the following 15 minutes.
The calibration is performed in multiple consecutive intervals organized in a pipeline with three steps:
\begin{itemize}
\item \textbf{Step A}: Incoming data is reconstructed using the last valid calibration (or the default calibration at the very beginning).
Based upon the reconstruction, new calibration is computed.
This is performed as long as needed such that sufficiently many events are processed to produce valid calibration.
\item \textbf{Step B}: The calibration result is propagated back to the beginning of the reconstruction chain such that it can be applied to the reconstruction.
Necessary postprocessing steps are performed as well such as the preparation of a new TPC cluster transformation map\footnote{The TPC cluster transformation map is needed to convert the raw TPC clusters from TPC pad row, TPC pad, and time coordinates into~$x$, $y$, and~$z$ coordinates taking into account the calibration.}.
\item \textbf{Step C}: The new calibration is now used for the reconstruction as long as it is valid, or until a new calibration is available.
\end{itemize}
Figure~\ref{fig:plot-calib-scheme} illustrates the online calibration scheme.
The steps are processed in a pipeline, such that while the calibration result is fed back (Step B, brown) and then used in the reconstruction (Step C, purple), a new calibration is computed in parallel (Step A, blue).
Accordingly, the reconstruction is not calibrated for events recorded at the very beginning of a run before Step B finishes.
The final reconstruction objects stored at the end of the process for offline use are valid for all events, also those at the beginning.
This scheme is feasible as long as the total time of steps A, B, and C is below the stability interval, e.\,g.~15 minutes in case of the TPC drift time calibration.

\begin{figure}[t]
\centering
\includegraphics[width=0.5\textwidth]{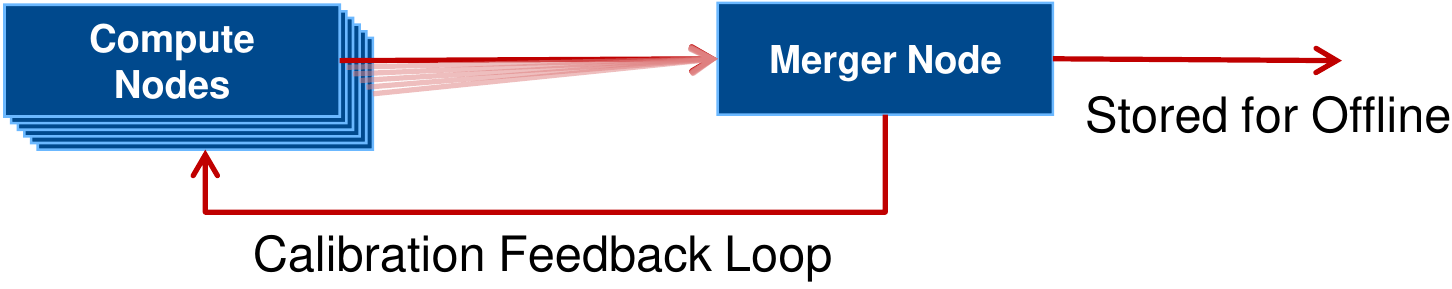}
\caption{Illustration of the merging of calibration objects.}
\label{fig:plot-calib-merger}
\end{figure}

Naturally, a single instance of the calibration software running on a single processor can not compute the calibration objects in time.
The HLT runs the calibration component on 172 of its 180 compute nodes with 3 instances of the calibration per compute node, for a total of 516 calibration processes that run in parallel.
These instances process the incoming events in a round robin fashion and they regularly send their calibration data to one single calibration merger process running on a dedicated calibration compute node.
This calibration node merges all the calibration data and creates the final calibration objects.
The objects are stored for offline use and in parallel shipped back to all compute nodes by the feedback loop, such that the compute nodes can perform the reconstruction based on the new calibration.
Figure~\ref{fig:plot-calib-merger} illustrates the process.

\subsection{Requirements for online calibration}
\label{sec:Requirements for online calibration}

\looseness=-1
On top of the time constraints from the above approach, online calibration poses a couple of requirements on the reconstruction and data transport framework, which are discussed in the following.
This subsection concludes with a list of requirements that needed to be solved to run the TPC drift time calibration in the ALICE HLT.
The following sections show how ALICE deals with these challenges.

\looseness=-1
The custom data transport framework in the HLT can be seen as a directed graph.
The fibres from the detector are input nodes in the graph, the network connections to DAQ (Data Acquisition) are output nodes, and the remaining nodes in the graph are processing components.
The links in the graph define the data flow.
All processing components process the events in an event-synchronous way, i.\,e.~they process one event after another in a pipeline.
Long running tasks as needed for the calibration pose a problem, because they stall the pipeline during the processing of the event.
All processing components which are in the processing graph before the stalled process become stalled as well as soon as the buffers run full.
Another issue is that the HLT processing chain must be loop-free for technical reasons by design.
This stems from the event-synchronous approach.
A loop in the graph would impose a cyclic dependency, i.\,e.~two processing components will wait for each other to finish processing the same event.

The TPC drift time calibration matches tracks reconstructed in the TPC to tracks in the ITS.
This means the HLT must at least provide track reconstruction for TPC and for ITS.
On top of that, the reconstruction of the tracks for these detectors should be standalone, i.\,e.~independent, in order to exclude the introduction of a bias.
Although the ITS delivers significantly less data than the TPC, the tracking for ITS is compute-intensive due to combinatorics because the ITS sits in the center of ALICE where the track density is the highest.
Also, scattering in the silicon layers is more complicated than inside the gas volume of the TPC.
Finally, ITS tracks have only up to 6 hits compared to more than 100 in the TPC, which requires a much more robust seeding procedure in order to find all tracks.
Therefore, the default HLT approach for ITS tracking is to prolong TPC tracks into the ITS and then collect the ITS hits close to the extrapolated tracks.
This poses two problems:
first, it could introduce a bias to use prolonged TPC tracks for the calibration of the TPC.
Second, the prolongation into ITS where the track density is very high requires high precision and works well only if the TPC is calibrated.
This leads to a chicken and egg problem.

Normal physics analysis in ALICE is based on the ROOT analysis and statistic software package.
The reconstruction creates C++ ROOT objects (see Section~\ref{sec:esd}), which are then used by the calibration tasks.
One paradigm in the ALICE HLT for online calibration is to use the same code for the calibration tasks that is used offline.
This reduces code duplication and simplifies the verification of the calibration code.
Processing components in the HLT are individual processes which cannot exchange C++ objects via pointers because they to not share a common address space.
The standard approach is serialization and deserialization of the object which causes significant CPU load.
Hence, HLT software should use only flat data structures which can be shipped to other processes directly.
This is incompatible with the standard Offline reconstruction programming model.

Usually, calibration is performed relative to a default calibration.
For instance, the TPC measures the clusters in row, pad, and time coordinates which must be transformed into spacial coordinates in order to run the track reconstruction and then the calibration.
This transformation is performed using the initial, default calibration.
The calibration uses the transformed clusters, and the exact transformation that was used to obtain the clusters's spacial coordinates has an impact on the calibration output.
In other words, the calibration must run on clusters transformed according to the default calibration but not on calibrated clusters.

\textit{List of requirements for online TPC drift time calibration in the ALICE HLT:}
\begin{itemize}
\item Fast reconstruction algorithms for ITS and TPC.
\item Independent standalone tracking algorithms for ITS and TPC.
\item Support for long-running tasks in the HLT framework.
\item Support for loops in the HLT data flow.
\item Data structures enabling fast data exchange between processing components.
\item The feedback loop must apply the calibration only for the tracking, but not to the calibration component.
\item Calibration process and feedback loop must not take longer than the time during which the calibration remains stable.
\end{itemize}

\subsection{Framework improvements}

\looseness=-1
Adding the feedback loop directly to the loop-free HLT processing chain is a major architectural change.
We wanted to avoid this since the current HLT data transport is thoroughly tested and we preferred incremental changes.
In particular, we prefer adding additional processing or communication components over changing the basis of the framework.
Considering the general approach for online calibration in the HLT, the calibration does not need to be event-synchronous.
The calibration result is not fed back to the reconstruction of the same event but at some later point in time.
This can happen asynchronously.
The HLT uses additional source and sink components based on the ZeroMQ data transport library for new communication channels not foreseen in the original framework \cite[§5.3]{bib:chep2015}.

The processing rate in the HLT is usually between 1\,kHz and 3\,kHz.
It is impossible to run long-running calibration tasks that last many seconds at this rate due to limited CPU resources.
Hence, the calibration task has to run only for a subset of the events, which is totally sufficient to gather enough statistics for the calibration.
Still, a long-running task blocks the chain even if it processes only a fraction of the events.
The problem is that it stalls the processing of that event for too long such that it affects all the other processes in the chain.
This means, even if the average processing time of the events would be short enough because many events are skipped, single events that need much time already block the chain.
This deficit of the event-synchronous processing approach is overcome by complementing the HLT with asynchronous processing components, which spawn a subtask in an asynchronous individual process, and then continue the fast synchronous event processing.
The result of the subtask is then used as soon as it is available~\cite[§5.1]{bib:chep2015}.
In order to protect ALICE data taking from fatal errors in the calibration code, the asynchronous processing can optionally happen in a completely isolated operating system process.
In this way, a possible memory leak or segmentation violation does not interfere with data taking but only breaks the processing of the calibration for few events until the process is restarted.

\section{Track reconstruction in the TPC}

The HLT employs GPU-accelerated track reconstruction for the ALICE TPC that is based on a Cellular Automaton principle to build track seeds: short track candidate of around 5 to 10 hits.
Afterwards, it uses the Kalman filter for track fitting and track following~\cite{bib:tns, bib:chep2012}.
The HLT employed 64 NVIDIA Geforce GTX480 GPUs during LHC Run 1.
The new HLT farm for LHC Run 2 is now equipped with 180 AMD FirePro S9000 GPUs.
One major concern with the original GPU tracker code was that it was based exclusively on the NVIDIA CUDA framework and was thus vendor-dependent.
For Run 2, an OpenCL implementation was created, which uses the AMD OpenCL C++ kernel languages extensions.
The code is written in a generic way, such that the same source code can be used with CUDA, OpenCL, and also for the CPU~\cite[§4]{bib:chep2015}.
This gives the greatest possible flexibility for hardware selection and reduces the maintenance effort.

Parallelization is implemented such that during the Cellular Automaton phase, 1 GPU thread handles 1 TPC cluster.
During the track following, 1 GPU thread handles 1 TPC track.
This allows for simple and efficient parallel processing, as the threads can operate almost fully independently.
Considering the amount of TPC clusters and tracks in a typical Pb-Pb event as well as the number of threads a GPU executes concurrently, this scheme resulted in full GPU utilization at the time it was implemented, e.\,g. for the GTX480 GPUs during Run 1.
Naturally, pp events with much less tracks do not use the GPUs efficiently, which is no problem however, because the GPUs are fast enough for pp reconstruction anyway.
In the meantime, the number of threads a GPU needs to execute in parallel to achieve full performance has increased significantly.
The number of tracks increased only slightly when the LHC moved from 3.5\,TeV/Z to 6.5\,TeV/Z in Run 2.
Overall, we see that now even single central Pb-Pb events are unable to load the S9000 GPUs of Run 2 to the full extent.
Looking ahead to Run 3, this problem becomes even more severe because new GPUs will feature even more parallelism.

\looseness=-1
The maximum data rate the TPC can deliver to the HLT is defined by the number of optical links times the link speed.
The HLT was tested with data replay at this maximum possible speed and it was proven to be able to run the full TPC reconstruction still having some margin to run reconstruction for other detectors.
Thus, development of HLT TPC track finding for Run~2 is complete.
Now, the focus lies on testing improvements for the Run~3 online computing facility already in the HLT during Run~2.

As first step, we increase the parallelism by processing multiple events on one GPU concurrently.
While during the implementation of the first GPU tracker version, GPUs could only execute one kernel at a time, modern GPUs can execute many independent kernels, even from different host applications, in parallel.
The HLT can run several instances of the tracker on one GPU processing multiple events concurrently, as long as there is enough GPU memory, which in case of Run 2 is sufficient for 3 central Pb-Pb events with pile up.
Table~\ref{tab:tpc-tracking} shows a first result.
Naturally, the wall time for a single event increases, which is not relevant.
(For reference, the time between a central Pb-Pb event reaches the HLT input until it leaves the HLT is in average around 3 seconds.)
In contrast, the processing throughput increases by~$31.8$\,\%.
Using all compute nodes at full capacity, the HLT can reconstruct~40,000,000 TPC tracks per second.

\begin{table}[htb]
\centering
\caption{Tracking time of HLT TPC GPU tracking processing two events in parallel (average over a selection of central Pb-Pb events)}
\label{tab:tpc-tracking}
\begin{tabular}{llr}
\hline
Number of events & Time & Time per event \\
\hline
1 & 145\,ms & 145\,ms \\
2 & 220\,ms & 110\,ms \\
\hline
\end{tabular}
\end{table}

\looseness=-1
Fortunately, the foreseen TPC readout scheme for Run 3 plays into our hands.
ALICE plans a TPC upgrade with continuous readout.
The online computing facility will no longer process single events but time frames with many overlapping events.
This will offer enough parallelism to use the GPUs to the full extent.
It is not clear yet whether the GPU memory will be sufficient to process the track finding of an entire time frame at once.
One solution is to slice the time frames along the time / beam axis, process the slices individually, and merge the track segments afterwards.
The feasibility of this approach is already shown by the track reconstruction in the current HLT, which processes the TPC sectors individually but concurrently, and merges the tracks segments afterward~\cite{bib:tns}.

Another foreseen development for Run 3 is the porting of additional online reconstruction components on the GPU, in particular, because modern GPUs can execute multiple kernels at the same time.
Canonical candidates for GPU adaptation are the ones before and after the GPU track finding in the HLT reconstruction chain (see Fig.~\ref{fig:plot-hlt-overview} in the last section).
As a first prototype, the final track refit, a substep of the TPC track merger and track fit component that merges the track segments reconstructed by the GPU track finding was ported to GPUs.
The prototype needs in average~6.8\,ms per event for the refit compared to 125.5\,ms on a single CPU Core (Intel Nehalem 2.8 GHz, the same events as in Table~\ref{tab:tpc-tracking}).
The bottleneck in this case is the PCI Express transfer, which takes longer than the computation itself.
Therefore, the most reasonable approach is to bring multiple successive components of the HLT chain onto the GPU, such that the data must not be transferred forth an back in between.

We plan to implement every new GPU processing component using a generic shared source code for CPU and GPU - in the same way as for the TPC tracking.
In addition, the implementations should be flexible enough to be applicable in the new software framework for online computing in Run 3 but also in the current HLT framework.
This will allow us to test new developments and benefit from them already in Run 2.

\section{Track reconstruction in the ITS}

\subsection{Scheme of ITS tracking in the HLT}

\begin{figure}[htb]
\centering
\includegraphics[width=0.8\textwidth]{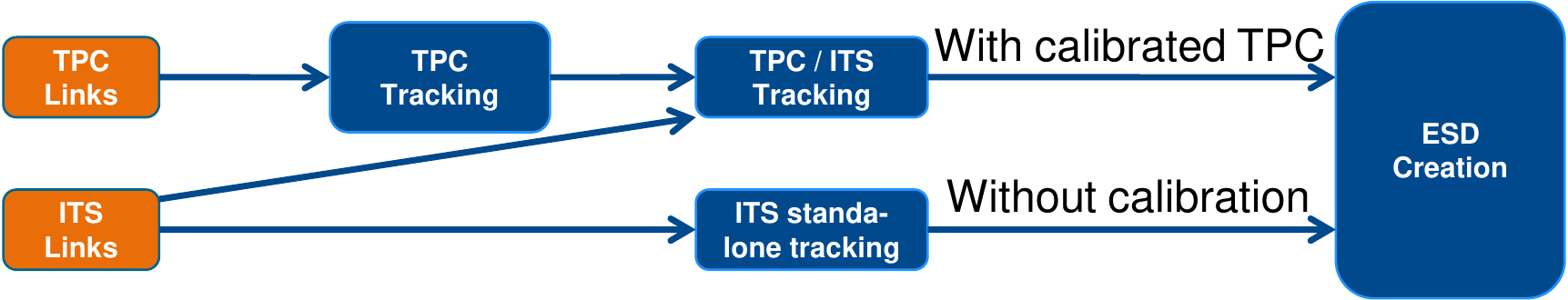}
\caption{Approach for ITS tracking in the HLT with two branches.
The lower branch runs the ITS standalone tracer providing ITS tracks in any case.
The upper branch uses TPC to ITS prolongation resulting in a higher efficiency, but it needs a calibrated TPC in order to achieve reasonable resolution.}
\label{fig:plot-its-tracker}
\end{figure}

The initial ITS tracking in the HLT which starts from prolonged TPC tracks is unsuited for online calibration.
Conversely, a full standalone ITS tracking has to deal with the excessive combinatorics inside ITS and would need too many CPU resources.
The HLT thus employs a hybrid approach with two independent ITS tracking branches.
The first is the traditional chain with prolonged TPC-ITS tracks.
In order to ensure good tracking results, the TPC needs to be calibrated.
In parallel, a second chain performs fast ITS tracking.
This is a fast standalone ITS tracking with some limitations.
In particular, it is not required to have maximum efficiency, i.\,e.~it does not need to find all tracks.
It only needs to find sufficiently many tracks for the ITS TPC matching in the calibration and for the luminous region estimation (see Section~\ref{sec:lumi}).
The scheme is visualized in Fig.~\ref{fig:plot-its-tracker}.
The following subsection describes the ITS standalone tracking in the HLT.

\subsection{Fast ITS standalone tracking}

The aim of the online ITS standalone reconstruction is to provide a sample of primary ITS tracks sufficient for calibration, without attempting to maximize the track-finding efficiency.
Instead, the emphasis is made on the processing speed and correctness of the tracking (minimization of random clusters attached to tracks).
The algorithm uses as an input the ITS clusters from all 6 layers and the primary vertex provided by the HLT components running upstream.
The latter is obtained as a position to which the maximum number of vectors connecting the two innermost ITS layers (silicon pixel detectors, SPD) converge.
The track reconstruction starts by rebuilding these vectors (tracklets), i.\,e.~finding pairs of SPD clusters seen under nearly the same angle from the vertex point.
The procedure is the optimized version of the off-line tracklet finding described in \cite{PRL2010}.

At the following step the tracks are found by following the tracklets to ITS layers at larger radii, where the ITS Silicon Drift and Strip Detectors (SDD and SSD, respectively) are .
For every tracklet a Kalman filter is initialized by the momentum estimated from the vertex and a pair of SPD clusters.
It is propagated outwards starting from the vertex, considered as a measured point.
The Kalman prediction/update is done first with already attached SPD clusters, then on the SDD and SSD clusters closest to extrapolation point, provided they meet a strict track-to-cluster~$\chi^{2}$ cut.
The magnetic field is taken to be a constant solenoidal one (approximation correct to~$\sim 10^{-3}$ in the ITS volume) while the multiple scattering is accounted for using average material budget per layer.
At least~4 clusters are required per track in the reconstruction in standard layout with all 6 ITS layers present and tracks with~2 consecutive layers without contribution are rejected.
Once the outermost layer is reached, the track {\it outward} kinematics is recorded (for the further matching with TPC) and inward
Kalman propagation is performed to obtain the kinematics at the vertex region.

The benchmark with simulated data (on a single core of i7-2600 CPU @ 3.40GHz) shows in $pp$ and $p-Pb$ events more than $2~kHz$ processing rate with a reconstruction efficiency (with respect to reconstructable Monte Carlo tracks) exceeding $90\%$ at $p_{T}>300 MeV/c$ and {\it fake } tracks contamination staying below $3\%$.
Minimum bias $Pb-Pb$ events are reconstructed at $18Hz$ rate ($40Hz$ skipping $15\%$ of most central collisions).
The efficiency above $p_{T}>300 MeV/c$ drops to $\sim~85\%$ and the fake tracks contamination is below $10\%$.

\subsection{Luminous region estimation based on ITS tracking}

\label{sec:lumi}

\looseness=-1
The volume where the particle beams overlap and most interactions take place is called the luminous region (LR).
It can only be reliably determined by the experiment itself by means of reconstructing and localizing each interaction and statistically determining the size and shape of the beam overlap volume.
This measurement, performed in (quasi-) real time, is used by the LHC to optimize the beam parameters.
The computation of LR requires accurate tracking close to the interaction point.
TPC tracks extrapolated~$\approx 80$\,cm towards the vertex have insufficient accuracy in this respect if the TPC is not fully calibrated.
A more robust method is to use ITS standalone tracking which does not require large extrapolation steps or the same degree of time dependent alignment and calibration as the TPC. The ITS standalone tracker has sufficient tracking efficiency and resolution.
This method is successfully implemented in the HLT and provides real time LR information to the LHC.

\looseness=-1
The vertex determination is implemented in the same component which performs the ITS standalone tracking.
First the tracks are propagated to the beam-line, then a fast linear fitter is deployed to find a vertex as a point minimizing the distance of the closest approach for maximum number of tracks.
The outlier tracks rejection is achieved by means of bi-square weighting filter, as described in \cite{IJMP2014}.

\section{Data structures for fast data transport in the HLT}

\label{sec:esd}

\subsection{A flat data structure}
As discussed in~\ref{sec:Requirements for online calibration}, one of the crucial features of the calibration procedures running in the HLT is the fact that they use the same code as used in the offline calibration.
This was developed to work on the output of the offline reconstruction, i.\,e.~on the C++ structures called Event Summary Data (ESD) on which the ALICE analyses are based.
The ESDs are not suited to be used in the HLT framework since the multiple processes that may run in the HLT do not share the same address space.
This implies that shipping C++ objects (possibly in the form of ROOT objects) between HLT processes can be done only through their serialization and deserialization, which introduce an unacceptable overhead in the framework for every process that would access the ESDs.
To overcome such difficulty when wanting to run in the HLT any task that would take as input the standard ESDs when running it offline (e.\,g.~calibration tasks, Quality Assurance, analysis...), the output of the HLT reconstruction is stored in flat structures, which are exchanged between the different HLT reconstruction, calibration, and QA components complying with the HLT framework requirements.

The development and implementation of the flat ESDs is based on the C++ concepts of inheritance and polymorphism.
A common base class for the flat ESDs and the standard ESDs allows to run the same calibration algorithm online in the HLT and offline.

\begin{figure}[b]
\centering
\includegraphics[width=13cm]{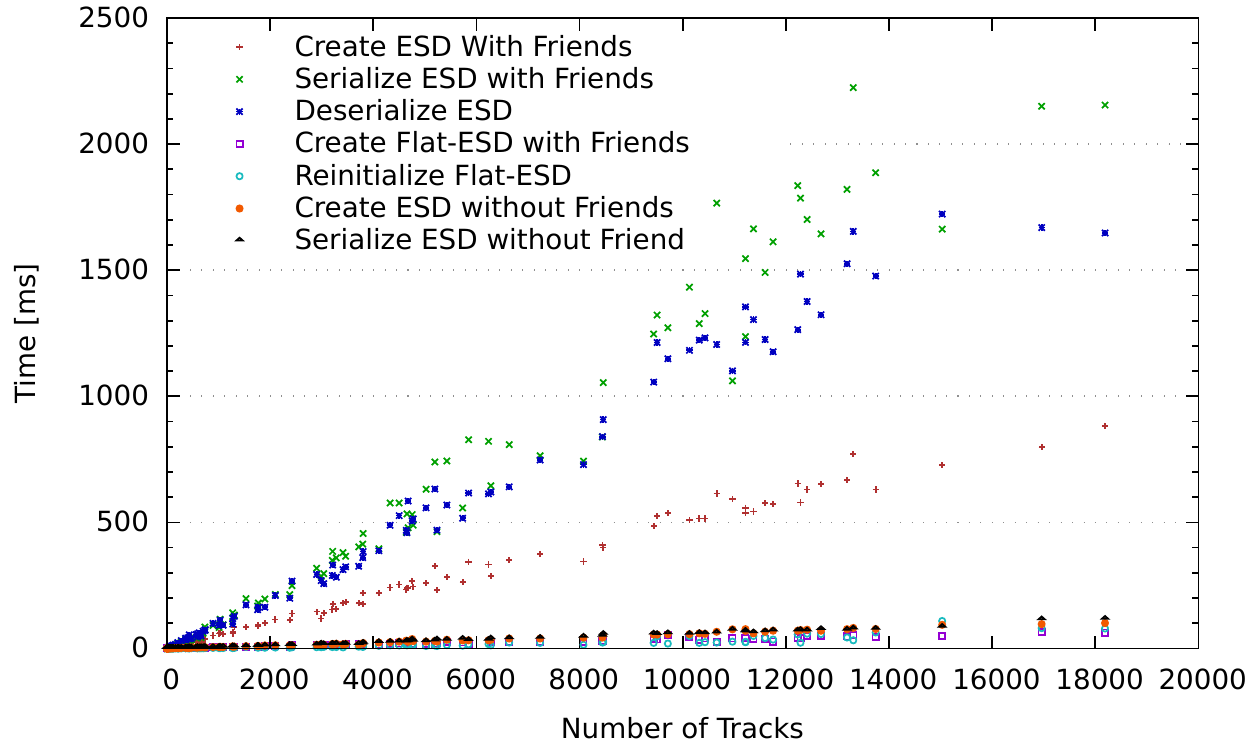}
\caption{Time needed for different steps in the creation and manipulation of the standard and flat ESDs as a function of the track multiplicity of the event, obtained processing Pb-Pb (from 2015) events.}
\label{fig:gp_flat_esd}       
\end{figure}

\begin{figure}[b]
\centering
\includegraphics[width=13cm]{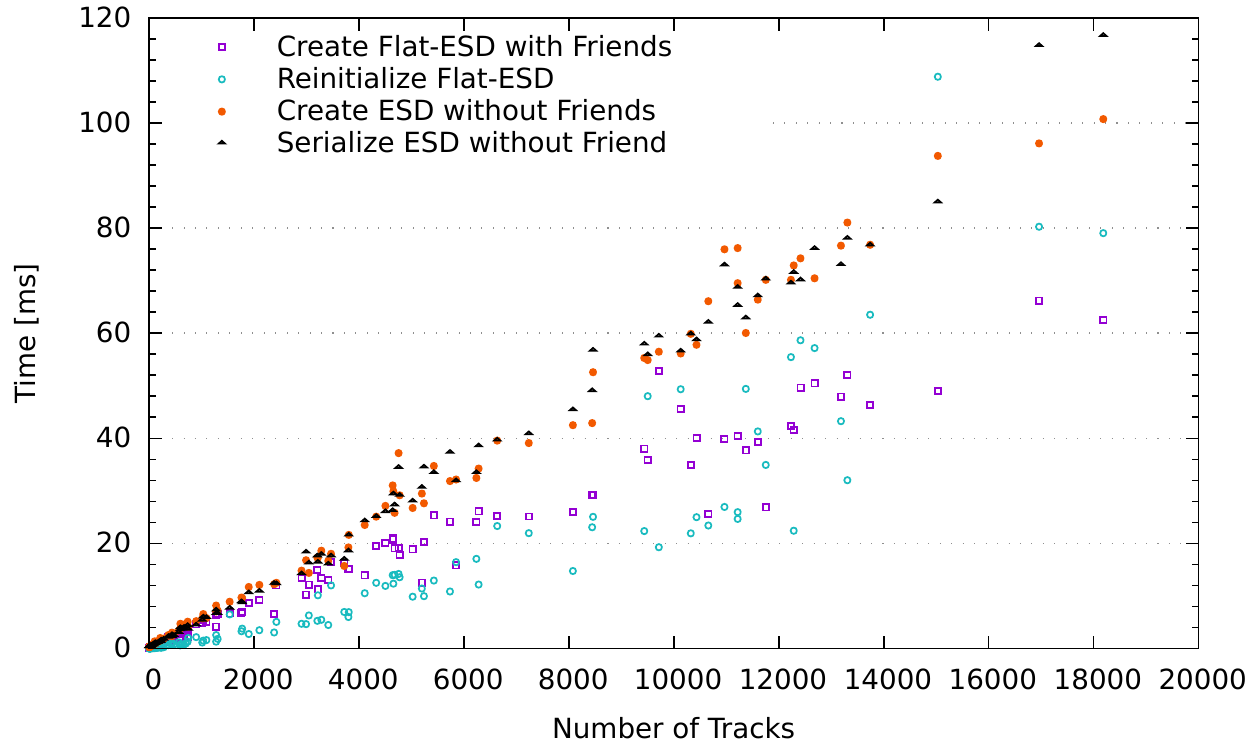}
\caption{Time needed for different steps in the creation and manipulation of the standard and flat ESDs as a function of the track multiplicity of the event.}
\label{fig:gp_flat_esd_no_friends}       
\end{figure}

While in the ESDs the different objects that come out of the reconstruction are stored inside the ESDs as different C++ objects, in the flat ESDs they are simply stored consecutively in memory and the bookkeeping of their position in the object is used to access them.
One special case among these objects is the so-called ESD friends, which is an object meant to store information not needed for analysis, but specifically for calibration.
For example, the information about the clusters used to form the tracks are there.
Since every track can have at maximum 159 clusters associated in the TPC (due to the same number of pad-rows in the detector), the amount of data in the friends is very large, and the size of the friends can be several times the one of the ESD tracks.

\subsection{Benchmarks}

\looseness=-1
We have evaluated the performance of the flat structures compared to standard ESDs with several benchmarks.
These are based on the output of the HLT reconstruction when stored in the standard ESD objects, or in the flat structures.
Figure~\ref{fig:gp_flat_esd} shows the time needed to create and manipulate (serialize, deserialize) the standard ESDs and the flat ESDs as a function of the track multiplicity of an event (based on a sample of around~500 Pb-Pb events taken in~2015).
As one can see, the time for the different stages grows linearly with the track multiplicity.
The time needed to create the flat ESDs including the friends is $\sim$10 times smaller than the time needed to create the standard ESDs with friends.
The serialization (and deserialization) of the standard ESDs including friends is between 2 and~3 times (and~$\sim 1.5$) times slower than their creation.
The reinitialization of the flat ESDs is a step needed in order to restore the virtual table of the flat ESD object, which is needed for the common interface of ESD and flat ESD.
The plot also shows the case when the standard ESDs are created and serialized without the friends, although not appropriate for calibration purposes.
The time in this case is of the same order as the one needed to create the flat ESDs with the full calibration information.
The behaviour of the standard ESDs without the friends compared to the flat structure is better visible in Fig.~\ref{fig:gp_flat_esd_no_friends}: also in this case the flat structures with the complete calibration information are more performant.

\begin{figure}[t]
\centering
\includegraphics[width=13cm]{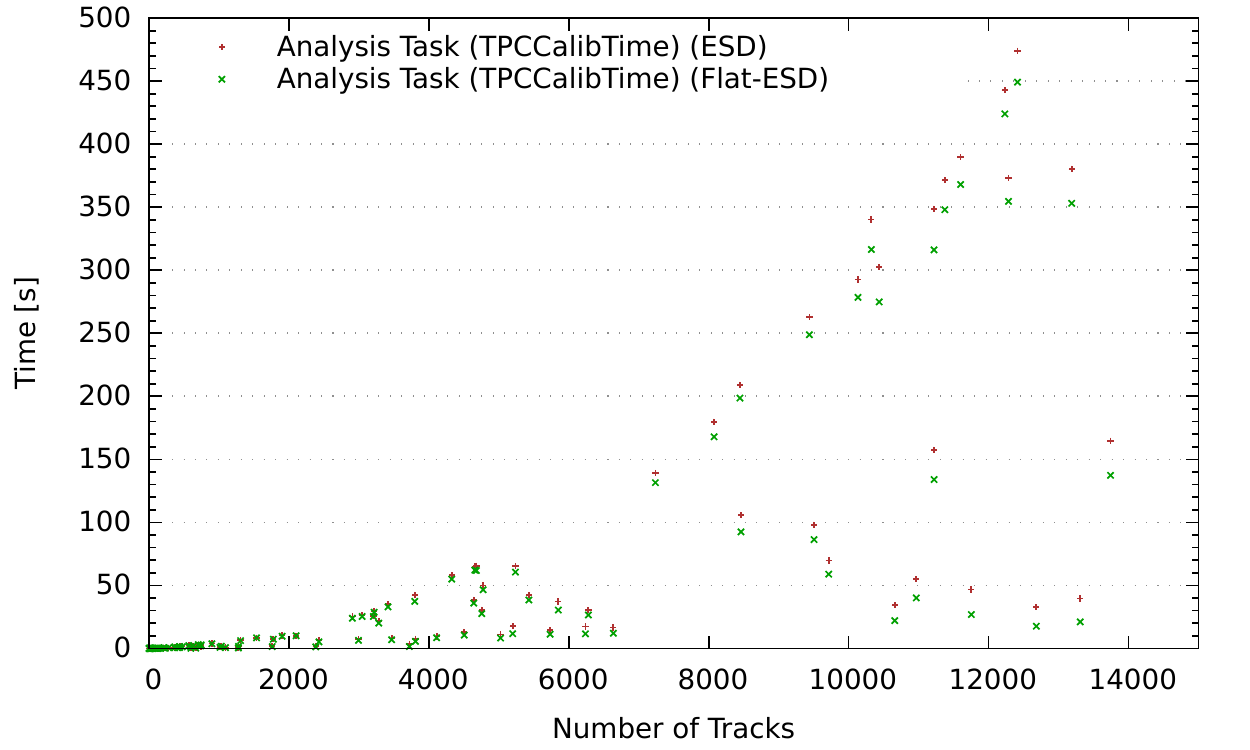}
\caption{Time needed for different steps in the creation and manipulation of the standard (without friends) and flat ESDs as a function of the track multiplicity of the event, obtained processing Pb-Pb (from 2015) events.}
\label{fig:gp_flat_esd_task}       
\end{figure}

\looseness=-1
Figure~\ref{fig:gp_flat_esd_task} shows the time needed by the TPC drift time calibration task when running on standard ESDs and on flat ESDs.
As one can see, the task performance is comparable when using as input the standard ESDs or the flat ones, being a few percent faster in case of the flat ESDs at high multiplicities.
This is an additional minor advantage of the flat ESDs.
The main advantage of the flat structures lies in the speed with which they are created and can then be shipped between different HLT components, more than in how fast they can be processed by the calibration task.
This is anyway running asynchronously with respect to data taking, and as such it does not impose any performance limitation.

\looseness=-1
Table~\ref{tab-esd-cpu} summarizes the resources needed to create the standard ESDs and the flat ESDs (with and without friends) in the HLT in number of CPU cores.
Considering the total amount of resources available in the HLT cluster, i.\,e.~$\sim 2000$ CPU cores, one realizes immediately that while the creation of standard ESDs as natural output of the HLT reconstruction is an expensive but well affordable process, adding the extra information needed to perform calibration tasks (i.\,e.~what is stored in the friends) would require half of the total HLT resources, which is completely infeasible.
In opposite, the flat ESDs can be created in the HLT together with the extra calibration data using a limited amount of resource, which is even smaller than those needed to generate the standard ESDs without friends.

\begin{table}[htb]
\centering
\caption{Number of CPU cores required to create normal and flat ESDs with and without friends obtained processing Pb-Pb (from 2015) events at 300 Hz. }
\label{tab-esd-cpu}
\begin{tabular}{llr}
\hline
ESD Type & with / without friends & Number of CPU cores \\
\hline
ESD Converter & without friends & 94.7 \\
ESD Converter & with friends & 947.0 \\
Flat-ESD Converter & without friends & 11.0 \\
Flat-ESD Converter & with friends & 72.4 \\
\hline
\end{tabular}
\end{table}

\section{Overview of HLT online calibration and a first online calibration test run}

\begin{figure}[b]
\centering
\includegraphics[width=0.98\textwidth]{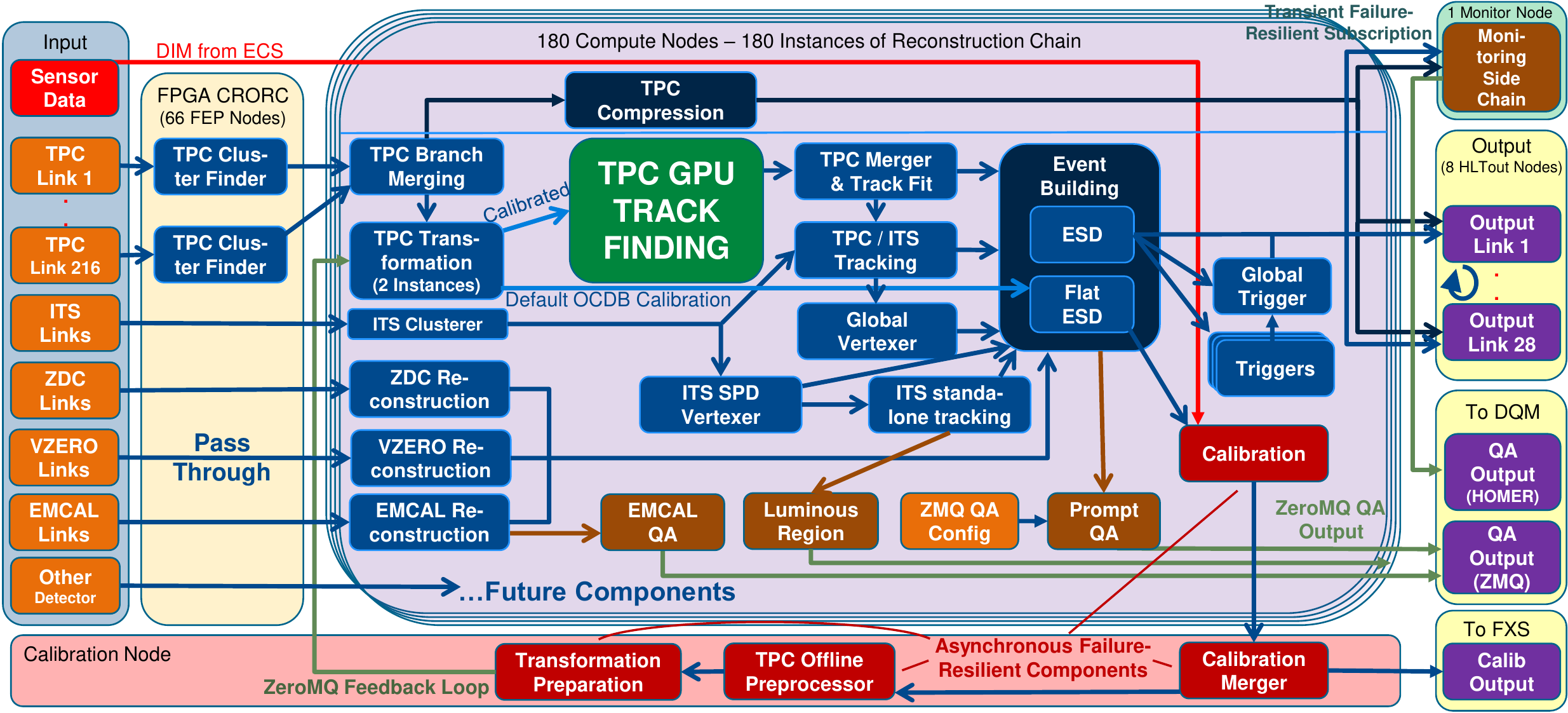}
\caption{Overview of all processing components and data flow in the HLT.}
\label{fig:plot-hlt-overview}
\end{figure}

\looseness=-1
Figure~\ref{fig:plot-hlt-overview} gives an overview over the data flow and all components in the HLT.
The actual components used for online calibration are shown in red, and they use the new asynchronous processing feature.
The actual calibration component is the only one that runs with many instances on all the compute nodes.
There is only one instance of the merger and the preprocessors, running on a dedicated calibration node.
Then, the new TPC transformation maps based on the new calibration is distributed in the cluster via ZeroMQ.
In order to make sure that the calibration component does not run on calibrated clusters, the TPC transformation component runs two instances.
One uses the new transformation map from the feedback loop.
The clusters from this instance are shipped to the tracker component.
The other instance uses always the default transformation map and ships the clusters to the ESD, from where they go to the calibration component.
The calibrated clusters in the tracks and the uncalibrated clusters for the ESD friends are identified via the cluster index which is unchanged by the cluster transformation.
Thus, the feedback loop improves the track finding, but does not interfere with the calibration itself.

Therefore, all the requirements of online calibration in the HLT listed in Section~\ref{sec:Requirements for online calibration} are met.
A first test of the above-described new features with online calibration was performed in December 2015.
The calibration components were running online under real conditions during Pb-Pb data taking in ALICE at LHC design luminosity.
This test proved that the HLT can handle the online calibration in a high load scenario with the highest data rates.

\looseness=-1
A first analysis of the processing rate of the calibration component shows the following:
the processing time of the calibration task for individual events can be quite long.
In particular, it depends superlinearly on the number of tracks in the event because of the TPC ITS matching.
During this first test, it took up to 15 minutes for the largest events, which is too long in order to use the results in the feedback loop.
Fortunately, the fraction of events which need more than 5 minutes for the calibration is only 2\,\%.
In particular, it is not necessary to run the calibration for the large events at all.
It is essential to have enough tracks for the TPC ITS matching in total.
It is faster to process the same number of tracks in many smaller events compared to fewer large events.
Excluding the small fraction of events that take very long, the processing of 5000 events (Step A), which is considered sufficient for the TPC drift time calibration in Pb-Pb, takes less than 5 minutes.
The feedback loop (Step B) takes around 20 seconds.
If the calibration is used for 5 minutes afterwards (Step C), the total time is below the limit of 15 minutes where the calibration is stable.
On top of this, this first test used the plain offline calibration software.
For the future, we plan to apply code optimizations to speed up the calibration task.

\end{document}